\documentstyle[epsfig,rotating,12pt]{article}

\catcode`\@=11
\long\def\@makefntext#1{
\protect\noindent \hbox to 3.2pt {\hskip-.9pt  
$^{{\ninerm\@thefnmark}}$\hfil}#1\hfill}		%CAN BE USED 

\def\@makefnmark{\hbox to 0pt{$^{\@thefnmark}$\hss}}  %ORIGINAL 
	
\def\ps@myheadings{\let\@mkboth\@gobbletwo
\def\@oddhead{\hbox{}
\rightmark\hfil\ninerm\thepage}   
\def\@oddfoot{}\def\@evenhead{\ninerm\thepage\hfil
\leftmark\hbox{}}\def\@evenfoot{}
\def\sectionmark##1{}\def\subsectionmark##1{}}

\renewcommand{\thefootnote}{\fnsymbol{footnote}}

\newcounter{sectionc}\newcounter{subsectionc}\newcounter{subsubsectionc}
\renewcommand{\section}[1] {\vspace*{0.6cm}\addtocounter{sectionc}{1} 
\setcounter{subsectionc}{0}\setcounter{subsubsectionc}{0}\noindent 
	{\normalsize\bf\thesectionc. #1}\par\vspace*{0.4cm}}
\renewcommand{\subsection}[1] {\vspace*{0.6cm}\addtocounter{subsectionc}{1} 
	\setcounter{subsubsectionc}{0}\noindent 
	{\normalsize\it\thesectionc.\thesubsectionc. #1}\par\vspace*{0.4cm}}
\renewcommand{\subsubsection}[1]
{\vspace*{0.6cm}\addtocounter{subsubsectionc}{1}
	\noindent {\normalsize\rm\thesectionc.\thesubsectionc.\thesubsubsectionc. 
	#1}\par\vspace*{0.4cm}}

\newcounter{appendixc}
\newcounter{subappendixc}[appendixc]
\newcounter{subsubappendixc}[subappendixc]

\renewcommand{\appendix}[1] {\vspace*{0.6cm}
        \refstepcounter{appendixc}
        \setcounter{figure}{0}
        \setcounter{table}{0}
        \setcounter{equation}{0}
        \renewcommand{\thefigure}{\Alph{appendixc}.\arabic{figure}}
        \renewcommand{\thetable}{\Alph{appendixc}.\arabic{table}}
        \renewcommand{\theappendixc}{\Alph{appendixc}}
        \renewcommand{\theequation}{\Alph{appendixc}.\arabic{equation}}
        \noindent{\bf Appendix \theappendixc #1}\par\vspace*{0.4cm}}

\def\abstracts#1{{
	\centering{\begin{minipage}{12.2truecm}\footnotesize\baselineskip=12pt\noindent
	\centerline{\footnotesize ABSTRACT}\vspace*{0.3cm}
	\parindent=0pt #1
	\end{minipage}}\par}} 

\renewenvironment{thebibliography}[1]
	{\begin{list}{\arabic{enumi}.}
	{\usecounter{enumi}\setlength{\parsep}{0pt}
\setlength{\leftmargin 1.25cm}{\rightmargin 0pt}
	 \setlength{\itemsep}{0pt} \settowidth
	{\labelwidth}{#1.}\sloppy}}{\end{list}}

\newcounter{itemlistc}
\newcounter{romanlistc}
\newcounter{alphlistc}
\newcounter{arabiclistc}

\newcommand{\fcaption}[1]{
        \refstepcounter{figure}
        \setbox\@tempboxa = \hbox{\footnotesize Fig.~\thefigure. #1}
        \ifdim \wd\@tempboxa > 6in
           {\begin{center}
        \parbox{6in}{\footnotesize\baselineskip=12pt Fig.~\thefigure. #1}
            \end{center}}
        \else
             {\begin{center}
             {\footnotesize Fig.~\thefigure. #1}
              \end{center}}
        \fi}

\newcommand{\tcaption}[1]{
        \refstepcounter{table}
        \setbox\@tempboxa = \hbox{\footnotesize Table~\thetable. #1}
        \ifdim \wd\@tempboxa > 6in
           {\begin{center}
        \parbox{6in}{\footnotesize\baselineskip=12pt Table~\thetable. #1}
            \end{center}}
        \else
             {\begin{center}
             {\footnotesize Table~\thetable. #1}
              \end{center}}
        \fi}

\def\@citex[#1]#2{\if@filesw\immediate\write\@auxout
	{\string\citation{#2}}\fi
\def\@citea{}\@cite{\@for\@citeb:=#2\do
	{\@citea\def\@citea{,}\@ifundefined
	{b@\@citeb}{{\bf ?}\@warning
	{Citation `\@citeb' on page \thepage \space undefined}}
	{\csname b@\@citeb\endcsname}}}{#1}}

\newif\if@cghi
\def\cite{\@cghitrue\@ifnextchar [{\@tempswatrue
	\@citex}{\@tempswafalse\@citex[]}}
\def\citelow{\@cghifalse\@ifnextchar [{\@tempswatrue
	\@citex}{\@tempswafalse\@citex[]}}
\def\@cite#1#2{{$\null^{#1}$\if@tempswa\typeout
	{IJCGA warning: optional citation argument 
	ignored: `#2'} \fi}}

 1
 1
 1

\font\ninerm=cmr9

\newcommand{\n}{\hspace*{-2.5mm}}
\begin{document}
\centerline{DESY 97-053\hfill March 1997}
\vspace{0.6cm}
\centerline{\normalsize\bf SM HIGGS DECAY AND SCATTERING}
\baselineskip=16pt
\centerline{\normalsize\bf PROCESSES AT TWO LOOPS\footnote{
Talk given at the Ringberg Workshop on
{\it The Higgs Puzzle --- What Can We Learn from LEP2, LHC, NLC and
  FMC?}, Munich, Germany, 8-13 December 1996.}}
%\vfill
\vspace*{0.3cm}
\centerline{\footnotesize KURT RIESSELMANN}
\baselineskip=13pt
\centerline{\footnotesize\it DESY-IfH Zeuthen, Platanenallee 6}
\baselineskip=12pt
\centerline{\footnotesize\it  D-15738 Zeuthen, Germany}
\centerline{\footnotesize E-mail: kurtr@ifh.de}
\vspace*{0.5cm}
\baselineskip=13pt

%\vfill
\vspace*{0.1cm} 
\abstracts{This contribution reviews the latest results of 
  the perturbative calculations of heavy-Higgs two-loop amplitudes.  A
  comparison of perturbative results with nonperturbative lattice
  calculations is made, and the theoretical uncertainties of the lower
  and upper bound on the Standard Model Higgs mass are presented.}
\vspace*{-0.2cm}
\normalsize\baselineskip=15pt
\setcounter{footnote}{0}
\renewcommand{\thefootnote}{\alph{footnote}}
\section{Introduction}
Why carrying out two-loop calculations for amplitudes involving the
Standard Model Higgs boson? Although the Higgs particle has not yet
been observed there are many aspects which raise questions to be
investigated at two loops. Topics include:
\begin{itemize}
\item the Higgs-mass dependence of precision variables: How well can
  we constrain the SM Higgs mass $M_H$ from the measurement of the
  $\rho$ parameter?\\[-0.6cm]

\item the leading heavy-Higgs corrections in powers of $G_FM_H^2$: How
  large are these corrections? For which value of $M_H$ do these power
  series cease to converge?\\[-0.6cm]

\item the radiative corrections to one-loop induced Higgs production
  processes such as gluon fusion: How much do higher-order corrections
  enhance such amplitudes?\\[-0.6cm]

\item the breakdown of perturbation theory at high energies: For which
  range of energies can we make reliable predictions of scattering
  cross sections?  Does a summation exist to improve the perturbative
  character?\\[-0.6cm]

\item the renormalization-group evolution of the Higgs coupling: What
  is the maximum energy scale up to which the Standard Model could be
  valid?
\end{itemize}

The answers to these questions advance our current understanding of
the Standard Model Higgs sector. They provide input for experimental
search strategies, help determining the various Higgs properties, and
shed new light on theoretical aspects of the Higgs particle.

In addition, calculations in the Higgs sector (broken $\Phi^4$ theory)
have been the frontier of numerical methods of quantum field theory in
form of lattice calculations.  Today, perturbative two-loop
calculations can be used to:%

\begin{itemize}
\item check the convergence of the perturbative series towards lattice
  results in the case of not-so-heavy Higgs masses.
\item apply numerical methods to the calculation of Feynman diagrams.
\end{itemize}

In this article I summarize the current status of those items
involving heavy-Higgs electroweak corrections.  Information on other
aspects with regard to the topics given above can be found elsewhere
in these proceedings.~\cite{speakers}

\section{The Lagrangian for heavy-Higgs radiative corrections}
\label{basics}

Neglecting gauge and Yukawa couplings, the Lagrangian of the standard
model Higgs sector reduces to
\begin{equation}
{\cal L}_H =
{\textstyle\frac{1}{2}}\left(\partial_\mu\Phi\right)^\dagger
\left(\partial^\mu\Phi\right)-{\textstyle\frac{1}{4}}\lambda
\left(\Phi^\dagger\Phi\right)^2+
{\textstyle\frac{1}{2}}\mu^2\Phi^\dagger\Phi\,, \label{l0higgs}
\end{equation}
where
\begin{equation}
\Phi = \left(
\begin{array}{c}
        w_1+iw_2\\
        h+iz
\end{array}
\right)=\left(
\begin{array}{c}
        \sqrt{2}w^{+}\\
        h+iz
\end{array}
\right). \label{phi} 
\end{equation}
The doublet $\Phi$ has a nonzero expectation value $v$ in the physical 
vacuum.
To facilitate perturbative calculations, the field $h$ is expanded
around the physical vacuum, absorbing the vacuum expectation value by
the shift $h\rightarrow H+v$. Hence the field $H$ has zero vacuum
expectation value.  Rewriting Eq.~(\ref{l0higgs}), ${\cal L}_H$ takes 
the form
\begin{eqnarray}
{\cal L}_H &=& 
{\textstyle\frac{1}{2}}\partial_\mu{\bf w}\cdot\partial^\mu{\bf w}
+{\textstyle\frac{1}{2}}\partial_\mu H\,\partial^\mu H
-{\textstyle\frac{1}{2}}M_H^2 H^2 + {\cal L}_{3pt} +{\cal L}_{4pt}\,, 
\label{lagrhiggs}
\end{eqnarray}
with the three-point and four-point interactions of the fields given by
\begin{eqnarray}
{\cal L}_{3pt}&=&-\lambda v\left({\bf w}^2H + H^3\right)\,,\\
{\cal L}_{4pt}&=&
-{\textstyle\frac{1}{4}}\lambda\left({\bf w}^4+2{\bf w}^2H^2+H^4
\right) \,.
\label{l1higgs}
\end{eqnarray}
Here $\bf w$ is the SO(3) vector of Goldstone scalars,
$(w_1,\,w_2,\,w_3)$, with $w_3=z$. The tadpole term and an additive
constant have been dropped.  The $w^\pm$ and $z$ bosons are massless,
in agreement with the Goldstone theorem. The Higgs mass
$M_H$ and the Higgs quartic coupling $\lambda$ are related by
\begin{equation}
\lambda=M_H^2/2v^2=G_FM_H^2/\sqrt{2},\label{lambda}
\end{equation} 
where $G_F$ is the Fermi constant, and $v=2^{-1/4}G_F^{-1/2}=246$ GeV. 

The Lagrangian ${\cal L}_H$ is the starting point for carrying out
calculations using the equivalence
theorem~\cite{eqt1,eqtren,eqtyuk}.  Using power-counting
arguments it has been shown that radiative corrections to
$O((G_FM_H^2)^n)$ = $O(\lambda^n)$ can also be calculated with the aid
of ${\cal L}_H$, that is, without having to use the full SM
Lagrangian.~\cite{eqtren} The implementation of proper
renormalization conditions is however crucial.

Including the Yukawa couplings by adding the fermionic Lagrangian
${\cal L}_F$ to ${\cal L}_H$, the basic Lagrangian for the calculation
of $O((G_FM_H^2)^n(G_Fm_t^2)^m)$ corrections is obtained. For a heavy
Higgs particle these are the leading {\it and} subleading electroweak
corrections, and they can be calculated using {\it massless} Goldstone
bosons, hence simplifying their calculation greatly.  For a Higgs mass
of less than approximately 300 GeV those corrections are not
leading anymore:~\cite{eqtyuk} The contributions from gauge
couplings need to be taken into account using the full SM Lagrangian.

\section{The decay $H\rightarrow W^+W^-,\, ZZ$: radiative corrections in
  powers of $G_FM_H^2$}

Using the limit $M_H\gg M_W$ the leading corrections to the bosonic
decay of the Higgs have been calculated to two
loops.~\cite{ghinculov,frink} {\it A priori} it is unknown for which
value of $M_H$ the two-loop correction term $O(\lambda^2)$ will
dominate over the one-loop term of $O(\lambda)$. Since $\lambda\propto
M_H^2/(246\,\, {\rm GeV})^2$ the breakdown of perturbation theory
might occur for values of $M_H$ less than 1 TeV.

The calculations of two-loop corrections to $H\rightarrow
W^+W^-,\,ZZ$~\cite{ghinculov,frink} are pioneering work with regard to
the use of numerical methods in the evaluation of Feynman diagrams of
three-point functions. The work by Ghinculov~\cite{ghinculov} uses
analytic cancellation of all ultraviolet divergencies using
dimensional regularization.  Infrared singularities are regularized
using a small mass for the Goldstone bosons which is taking to zero in
the final result. The finite contributions of the Feynman diagrams are
obtained by numerical integration of Feynman-parameter integrals.  The
calculation by Frink {\it et al.}~\cite{frink} features massless
Goldstone bosons.  Both UV and IR divergent Feynman diagrams are
calculated analytically, including their finite contributions.  The
sum of these diagrams leads to the explicit cancellation of both types
of divergencies.  The non-divergent Feynman diagrams are calculated
using numerical integration in orthogonal momentum space components.
The two results~\cite{ghinculov,frink} for the two-loop coefficients
agree to $1.7 \times 10^{-3}$, indicating the reliability at which
these numerical methods operate. In particular:~\cite{frink} 
\begin{eqnarray}
\label{hwwwidth}
\Gamma(H&\n\rightarrow\n& ZZ,\;W^+W^-) \propto 
\lambda(M_H)\left(1
%+ 2.800\,952\, \frac{\lambda}{16\pi^2} 
+ 2.800\,\dots\frac{\lambda}{16\pi^2} 
+ 62.030\,8(86) \frac{\lambda^2}{(16\pi^2)^2}\right)\,.
\end{eqnarray}

It is interesting to compare the size of the coefficients with the
leading heavy-Higgs corrections calculated in the case of fermionic
Higgs decay.  Using numerical or analytical methods one
obtains~\cite{hff1,hff2,hff3}
\begin{eqnarray}
\Gamma(H&\n\rightarrow\n& f\bar f) \propto
g_f^2  \left(1
+ 2.117\,\dots\frac{\lambda}{16\pi^2}
%+ 2.117\,203\dots\frac{\lambda}{16\pi^2}
- 32.656\,\dots\frac{\lambda^2}{(16\pi^2)^2}\right)\,.
%- 32.656\,530\dots\frac{\lambda^2}{(16\pi^2)^2}\right)\,.
\label{hffwidth}
\end{eqnarray}
Comparing the last two equations we see that the coefficients of the
perturbative series are of similar size. This is not the case for
scattering processes; see below.

\begin{figure}[hbt]
\vspace*{13pt}
\begin{center}
\mbox{\hspace*{-4mm}
\begin{turn}{90}
\epsfig{file=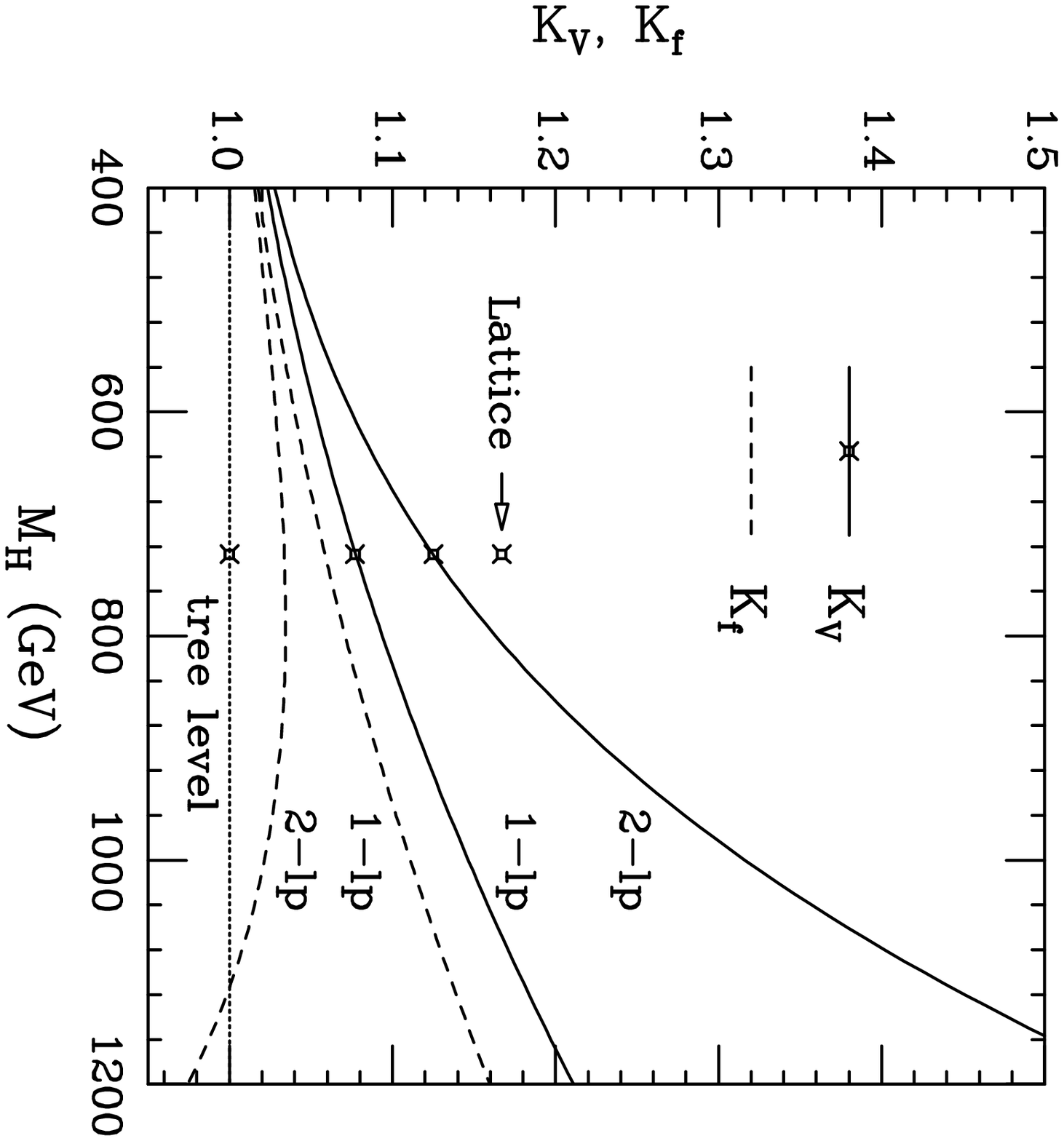,width=8.3cm}
\end{turn}}
%\vspace*{-10mm}
\end{center}
\fcaption{The $K$-factors for $H\rightarrow W^+W^-,\, ZZ$ and
  $H\rightarrow f\bar f$ at various orders in perturbation theory.
  For $M_H=727$ GeV the bosonic perturbative $K$-factor is compared
  with a result from lattice calculations.~\cite{goe}}
\label{fig:kfactors}
\end{figure}

The $K$-factors of both bosonic and fermionic decay width are given by
the expressions in the large brackets of Eq.~(\ref{hwwwidth}) and
(\ref{hffwidth}), respectively.  They are plotted in
Fig.~\ref{fig:kfactors} as a function of $M_H$.  The corrections are
less than 10\% for $M_H<670$ GeV, and for $M_H=980$ the corrections
are less than 30\%.  Yet perturbation theory is not meaningful for
large Higgs masses.  At $M_H=930$ GeV the two-loop bosonic correction
term is as large as the one-loop term. In the fermionic case the
two-loop correction term compensates the one-loop term if $M_H\approx
1100$ GeV.  A different criterion for judging the breakdown of
perturbation theory is the investigation of scale and scheme
dependence.~\cite{willey,nierste} For the Higgs decay processes this leads
to a perturbative bound on $M_H$ of about 700 GeV.~\cite{nierste}

In the case of bosonic Higgs decay, we are also able to compare the
perturbative results with nonperturbative computations carried out
using lattice techniques.~\cite{LW,goe} The lattice result
obtained~\cite{goe} for $M_H=727$ GeV appears to be consistent with
the perturbative results for $H\rightarrow W^+W^-$; see
Fig.~\ref{fig:kfactors}.  The difference between the two-loop
perturbative and the nonperturbative result can probably be
contributed to the missing higher-order perturbative correction terms
and the use of massive instead of massless Goldstone bosons (pions) in
the lattice calculation.  A detailed comparison of these results is in
preparation.

For completeness we also mention that the two-loop heavy-Higgs
correction to the loop-induced process $H\rightarrow \gamma \gamma$
has also been calculated.~\cite{koe} 

\section{Radiative corrections to scattering processes}

Subprocesses such as $W^+W^-\rightarrow W^+W^-$ are important to
extract experimental information on the Higgs resonance.  The
tree-level $O(\lambda\: \n\propto\n \: G_FM_H^2)$ contribution to this
cross section is entirely due to the scattering of longitudinally
polarized bosons, $W_L^+W_L^-\rightarrow W_L^+W_L^-$. In the case of a
heavy Higgs, this channel gives the dominant contribution to the cross
section. The transverse polarizations only couple via gauge couplings
and are suppressed as $g^2/\lambda\propto M_W^2/M_H^2$. In the case of
$ZZ\rightarrow ZZ$, however, it has been shown that radiative gauge
corrections can enhance the transverse channels
significantly.~\cite{denner} This is also expected to happen for
$W^+W^-\rightarrow W^+W^-$.

The dominant heavy-Higgs corrections to longitudinal scattering
amplitudes involving $Z_L$, $W_L$ or $H$ are known up to two
loops.~\cite{maher,rie} In contrast to the Higgs decay amplitudes
which only depend on one parameter (the coupling $\lambda$, or
equivalently, the Higgs mass $M_H$), the $2\rightarrow 2$ boson
scattering amplitudes also depend on the center-of-mass energy
$\sqrt{s}$ of the scattering process. In the high-energy limit terms
of order $M_H^2/s$ can be neglected, and the scattering amplitude
exhibits a purely logarithmic energy dependence.  The cross section
for $W_L^+W_L^-\rightarrow W_L^+W_L^-$ in on-mass-shell (OMS)
renormalization is~\cite{maher,rie}
\begin{eqnarray}
\label{wwwwcross}
\sigma(s)\, 
& = & \frac{1}{\pi s} \lambda^2\,
\Biggl[\,1\,+
\left( 24 \ln \frac{s}{M_H^2} -
       \, 48.64 \right) \,\frac{\lambda}{16\pi^2}\,\Biggr.\,
 \\ 
&& \phantom{\frac{1}{8\pi s} \lambda^2\, }
+ \left( 432 \ln^2 \frac{s}{M_H^2} - 2039.3\ln \frac{s}{M_H^2} 
   +\;3321.7\,\right) \frac{\lambda^2}{(16\pi^2)^2}\,.\nonumber\\
&& \phantom{\frac{1}{8\pi s} \lambda^2\, }
+\Biggl.\; {\rm O}\left(\lambda^3\right)
+\; {\rm O}\left(\frac{M_H^2}{s}\right) \,\Biggr]
+\; {\rm O}\left(g^2\right)
\,.\nonumber
\end{eqnarray}
The coefficients found here are more than a factor 10 larger than 
the coefficients for the decay widths given in Eqs.~(\ref{hwwwidth}) and
(\ref{hffwidth}). 

%
%-----------------------------------------------------------------------
\begin{figure}[tbh]
\vspace*{-13pt}
\begin{center}
\mbox{\hspace*{-4mm}
\begin{turn}{90}
\epsfig{file=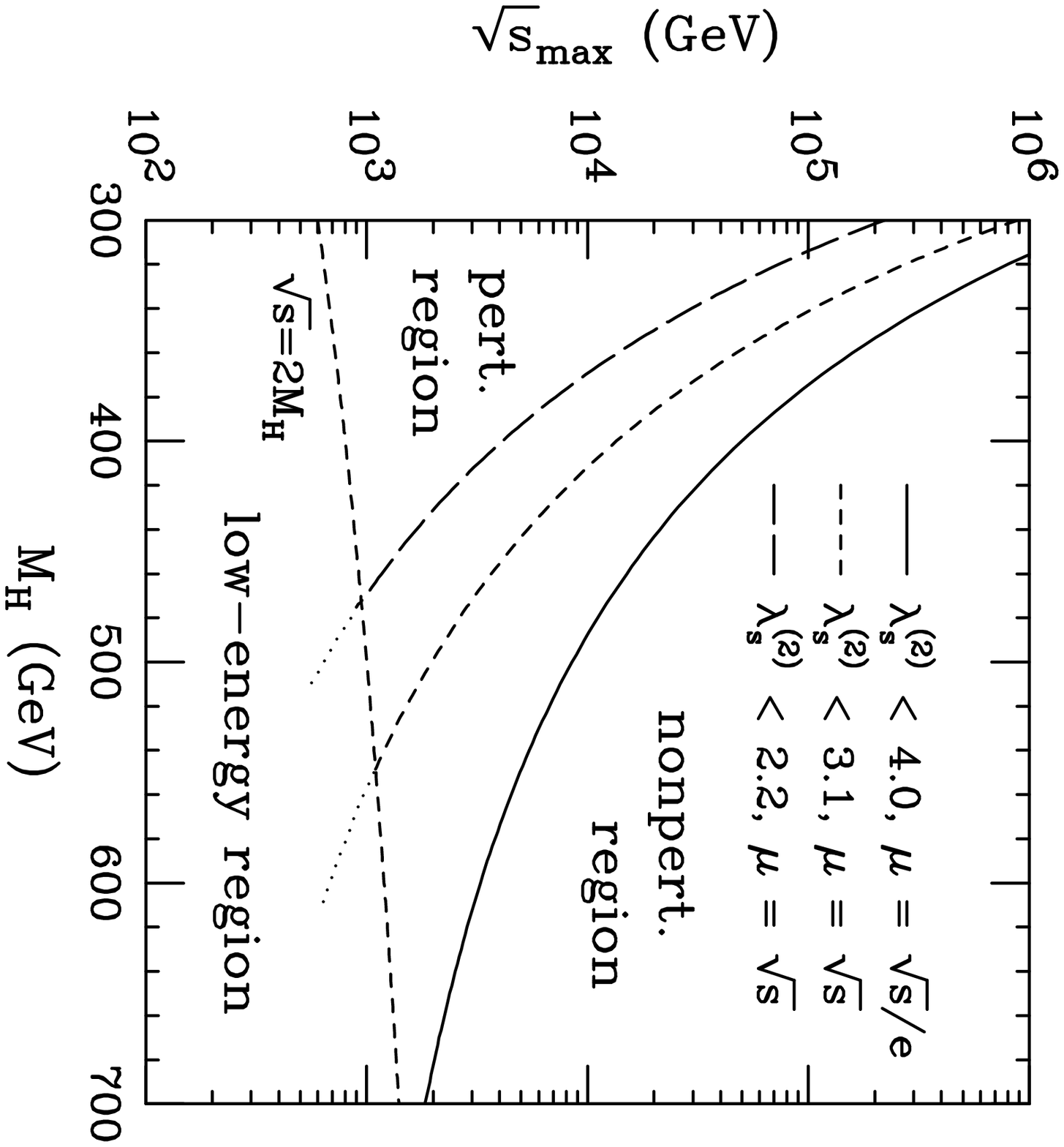,width=7.4cm}
\end{turn}}
%\vspace*{-10mm}
\end{center}
\fcaption{The perturbative limits on ($M_H$, $\protect\sqrt{s}$) found
  in high-energy scattering processes.  The limits on the running
  coupling are derived with~\cite{rw} (solid line) and
  without~\cite{joh,unit,rie} (dashed lines) summation.}
\label{fig:limits}
\end{figure}
%-----------------------------------------------------------------------
Applying renormalization-group methods, the logarithmic energy
dependence can be absorbed into a running Higgs coupling, which at one
loop is given by
\begin{eqnarray}
  \lambda(\mu)&=&{\lambda(M_H)} \left[1 -
    12\,\frac{\lambda(M_H)}{16\pi^2}\ln\left(\frac{\mu^2}{M_H^2}\right)
    \right]^{-1}\,,
\label{run1lp}
\end{eqnarray}
and $\lambda(M_H)$ is fixed by Eq.~(\ref{lambda}).  Using the
corresponding three-loop running coupling,~\cite{nierste}
Eq.~(\ref{wwwwcross}) can be rewritten as a
next-to-next-to-leading-logarithmic (NNLL) cross section,
\begin{eqnarray}
  \sigma\:\propto\:\frac{1}{s} \lambda^2(\sqrt{s})\left(1 - 48.64
  \frac{\lambda(\sqrt{s})}{16\pi^2} + 3321.7
  \frac{\lambda^2(\sqrt{s})}{(16\pi^2)^2}\right)\,.
\label{cross}
\end{eqnarray}
Here the natural choice $\mu=\sqrt{s}$ has been made, and a prefactor
with a small $s$-dependence due to non-zero anomalous dimensions has
been neglected.~\cite{rie}

It is striking that the one-loop cross section is {\it negative}
for the relatively low value  of
$\lambda(\sqrt{s}) \approx 3.2$. Using various criteria, the perturbative
limit on $\lambda$ can be given~\cite{nierste,rie} as
$\lambda(\sqrt{s}) \approx 2.2$.\footnote{Some authors use a
  different normalization of the Higgs potential, leading to a
  numerically different bound on the coupling. The bounds on the Higgs
  mass are unaffected by this redefinition.}  $\:$
Similar bounds on the running coupling are
obtained using arguments concerning unitarity
violations.~\cite{joh,unit} Tree-level unitarity bounds without the
use of the running coupling are independent of $\sqrt{s}$ and less
stringent:~\cite{dic} They require $\lambda=\lambda(M_H)< 4\pi/3
\approx 4.2$.

Recently it has been discovered that the approximate summation of a
subset of Feynman diagrams extends the range of validity of the
perturbative results.~\cite{rw} This summation corresponds to taking
$\mu=\sqrt{s}/e\approx \sqrt{s}/2.7$ as the appropriate choice of
scale in the running coupling.\footnote{Starting from the
  $\overline{\rm MS}$ scheme, this choice of $\mu$ leads to the
  $G$-scheme.~\cite{che}} $\:$
The high-energy NNLL cross section then reads
\begin{eqnarray}
\sigma\:\propto\:\frac{1}{s}
\lambda^2(\sqrt{s}/e)\left(1 - 0.64
\frac{\lambda(\sqrt{s}/e)}{16\pi^2} 
+ 923.1 \frac{\lambda^2(\sqrt{s}/e)}{(16\pi^2)^2}\right)\,.
\label{crossres}
\end{eqnarray}
The summed cross section is perturbative for much larger values of the
running coupling.  Using various criteria it has been
concluded~\cite{rw} that perturbative calculations in the Higgs sector
are reliable for a {\it running} Higgs coupling up to
$\lambda(\mu)\approx 4$, and perturbative unitarity is restored up to
this value.  This significantly extends the range in
$M_H$ and $\sqrt{s}$ for which high-energy calculations are
reliable; see Fig.~\ref{fig:limits}.\\

\section{Renormalization-group behaviour of $\lambda$ 
  including all SM couplings}

The one-loop running coupling introduced in the previous section,
Eq.~(\ref{run1lp}), is valid only if $M_H$ is large. Increasing the
scale $\mu$, the coupling increases monotonically, eventually
approaching the Landau singularity.  For small values of $M_H$ the
behaviour is different. In this case the contributions from gauge and
Yukawa couplings need to be included. In particular, the presence of
the top-quark Yukawa coupling $g_t$ can cause the Higgs running
coupling to decrease as $\mu$ increases, possibly leading to an
unphysical negative Higgs coupling. This is due to the negative
contribution of the top quark to the
one-loop beta function of the Higgs coupling:
\begin{equation}
\beta_\lambda=24\lambda^2+12\lambda g_t^2 - 6g_t^4 + {\rm gauge\:
  contributions}, 
\end{equation}
where all couplings must be taken to be running couplings. %

\begin{figure}[bht]
\vspace*{13pt}
\begin{center}
\mbox{\hspace*{-4mm}
\begin{turn}{90}
\epsfig{file=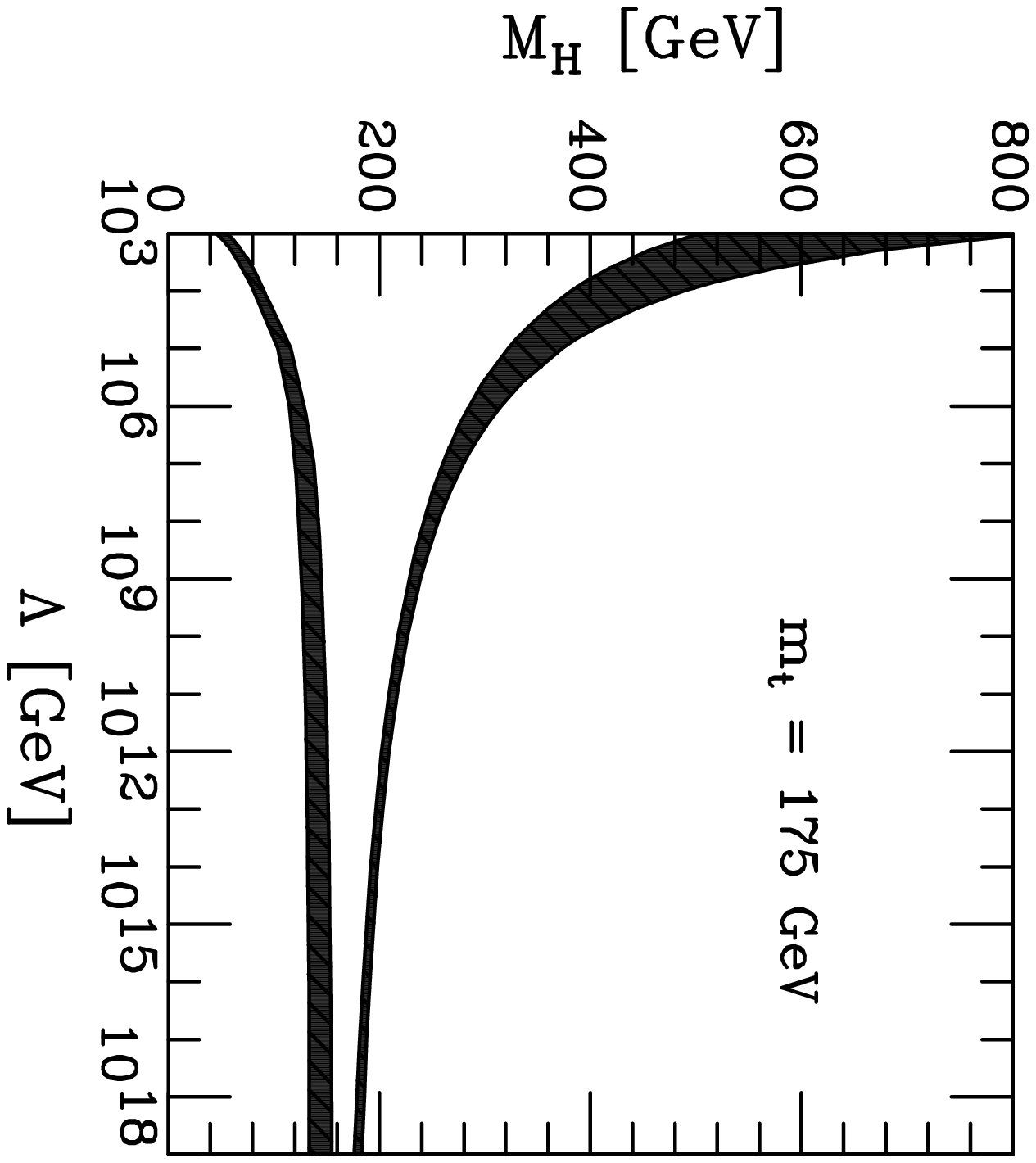,width=8.0cm}
\end{turn}}
%\vspace*{-10mm}
\end{center}
\fcaption{The present-day theoretical uncertainties on the
  lower~\cite{lower1,lower2} and upper~\cite{upper2} $M_H$ bounds when
  taking $m_t=175$ GeV and $\alpha_s(M_Z)=0.118$.}
\label{fig:lcplot2}
\end{figure}

Requiring the Higgs coupling to remain finite and positive up to an
energy scale $\Lambda$, constraints can be derived on the Higgs mass
$M_H$.~\cite{oldwork} Such analyses exist at the two-loop level for
both lower~\cite{lower1,lower2} and upper~\cite{upper1,upper2} Higgs
mass bounds.  Since all Standard Model parameters are experimentally
known except for the Higgs mass, the bound on $M_H$ can be plotted as
a function of the cutoff energy $\Lambda$.  Taking the top quark mass
to be 175 GeV and a QCD coupling $\alpha_s(M_Z)=0.118$ the result is
shown in Fig.~\ref{fig:lcplot2}.  %

The bands shown in Fig.~\ref{fig:lcplot2} indicate the theoretical
uncertainties due to various cutoff criteria, the inclusion of
matching conditions, and the choice of the matching
scale.~\cite{upper2} If the Higgs mass is 160 to 170 GeV then the
renormalization-group behaviour of the Standard Model is perturbative
and well-behaved up to the Planck scale $\Lambda_{Pl}\approx 10^{19}$
GeV.  For smaller or larger values of $M_H$ new physics must set in
below $\Lambda_{Pl}$.

%-----------------------------------------------------------------------

\section{Concluding remarks}

%About 50 physicists have participated in this workshop on ``The Higgs
%Puzzle''.  The strong believe in the existence of a
%Higgs boson can be read of the fact that about 65\% of
%the participants have already carried out two-loop investigations of
%Higgs renormalization-group properties or calculated decay- and
%scattering amplitudes involving the Higgs boson at two loops. 
 
The phenomenological aspects of a fundamental Higgs particle are well
understood, and it seems a matter of time and money to prove (or
disprove) its existence.  Finding such a Higgs particle, however,
would be just a first step. The multi-loop calculations presented in
this and other talks of the workshop will enable us to check many
properties of the Higgs boson, and comparison with experimental data
hopefully provides us with new insight for the development of a more
complete particle theory up to the Planck scale.

Since the experimental discovery of the Higgs boson at a collider
experiment may still take a long time, I would like to conclude with
my personal discovery of the Higgs Boson:
http://homepages.enterprise.net/hboson/homefiles/higgsb.html.

\end{document}

\bibitem{bij} A. Ghinculov and J.J. van der Bij,
Nucl.\ Phys.\ {\bf B436}  (1995) 30.

\end{document}